\documentclass[reprint,superscriptaddress,
amsmath,amssymb,aps,
prb,
]{revtex4-2}

\usepackage{pifont}
\usepackage{graphicx}
\usepackage{dcolumn}
\usepackage{bm}
\usepackage[hidelinks]{hyperref}
\usepackage[mathlines]{lineno}
\usepackage{color,soul}
\usepackage{float}

\begin{document}


\title{Identifying Electronic Doorway States\\in the Secondary Electron Emission from Layered Materials}



\author{A. Niggas}
\email{anna@iap.tuwien.ac.at}
\affiliation{Institute of Applied Physics, TU Wien, Vienna, Austria, EU}%

\author{M. Hao}%
\affiliation{Institute for Theoretical Physics, TU Wien, Vienna, Austria, EU}%

\author{P. Richter}
\affiliation{Institute of Physics, Chemnitz University of Technology, Chemnitz, Germany, EU}%
\affiliation{Center for Materials, Architectures and Integration of Nanomembranes (MAIN), Chemnitz University of Technology, Germany, EU}%

\author{F. Simperl}
\affiliation{Institute of Applied Physics, TU Wien, Vienna, Austria, EU}%

\author{F. Blödorn}
\affiliation{Institute of Applied Physics, TU Wien, Vienna, Austria, EU}%

\author{M. Cap}
\affiliation{Institute of Applied Physics, TU Wien, Vienna, Austria, EU}%

\author{J. Kero}
\affiliation{Institute of Applied Physics, TU Wien, Vienna, Austria, EU}%

\author{D. Hofmann}
\affiliation{Institute of Applied Physics, TU Wien, Vienna, Austria, EU}%

\author{A. Bellissimo}
\affiliation{Institute of Photonics, TU Wien, Vienna, Austria, EU}%

\author{J. Burgdörfer}
\affiliation{Institute for Theoretical Physics, TU Wien, Vienna, Austria, EU}%

\author{T. Seyller}
\affiliation{Institute of Physics, Chemnitz University of Technology, Chemnitz, Germany, EU}%
\affiliation{Center for Materials, Architectures and Integration of Nanomembranes (MAIN), Chemnitz University of Technology, Germany, EU}%

\author{R.A. Wilhelm}
\affiliation{Institute of Applied Physics, TU Wien, Vienna, Austria, EU}%

\author{F. Libisch}
\email{florian.libisch@tuwien.ac.at}
\affiliation{Institute for Theoretical Physics, TU Wien, Vienna, Austria, EU}%

\author{W.S.M. Werner}
\email{werner@iap.tuwien.ac.at}
\affiliation{Institute of Applied Physics, TU Wien, Vienna, Austria, EU}%

\date{\today}

\begin{abstract}
   We investigate the secondary low-energy electron emission induced by inelastic electron scattering from graphene and layered materials thereof. By applying a coincidence detection of the primary scattered and the emitted secondary electron we unravel pronounced resonance features otherwise overshadowed by the largely structureless secondary electron energy distribution. Supported by density functional theory calculations we show that these structures are the signature of prominent Feshbach resonances above the vacuum threshold which originate from interlayer states acting as a doorway state for electron emission. Remarkably, some of these doorway states open up only for samples with more than 5 layers.  
\end{abstract}

\maketitle


    Low-energy electron (LEE) emission plays a key role in today's high-resolution microscopy and nanotechnology.
    For example, the imaging contrast in scanning electron microscopy (SEM, Everhart-Thornley detectors)~\cite{amelinckx2008electron,seiler1983secondary,everhart1960wide} and helium ion microscopy (HIM)~\cite{bell2009contrast,ohya2009comparison,wirtz2019imaging} rely on secondary LEE emission. 
    Other LEE related techniques include electron-beam-induced chemical processing~\cite{arumainayagam2010low,bohler2013control}, deposition techniques (e.g. focused-electron-beam-induced deposition (FEBID))~\cite{thorman_role_2015,huth_focused_2012}) or LEE-driven damage in biomolecules~\cite{sanche2005low,sanche2002nanoscopic}, among many others. 
    Despite the importance of LEEs in a plethora of biological, chemical and physical processes a microscopic understanding of the underlying mechanisms which determine the energy profile of LEEs emitted from solid surfaces is still lacking. 

    One focus of the extensive literature on LEE emission~\cite{dekker1958secondary,scholtz1996secondary,shih1997secondary,bellissimo_secondary_2020} has been on carbon-based materials such as graphite~\cite{papagno_electronic_1983,moller_experimental_1982,willis_secondary-electron_1974,willis_secondary-electron_1974}. 
    Graphite is involved in many applications due to its low secondary electron yield, e.g. for electron optics and wall material in charged particle storage rings where multipacting is a well-known issue and needs to be reduced to a minimum~\cite{parodi2011multipacting,pinto2011thin}. 
    While many LEE spectra consist of a mainly featureless energy distribution, graphite stands out due to a distinct and non-dispersive feature at $\sim3.3$\,eV above the (sample) vacuum level $E\textsubscript{vac}$ (i.e. 7.9\,eV above the Fermi level $E\textsubscript{F}$)~\cite{PhysRevB.37.4482,werner_secondary_2020}. 
    This peak commonly referred to in the literature as the `X-peak'~\cite{yamane_low-energy_2001} originates from plasmon decay into single-particle hole excitations \cite{werner_secondary_2020}.
    Here we experimentally and theoretically demonstrate that such peaks in LEE spectra not only mirror structures in the density of states (DOS), but serve as unambiguous signatures of Feshbach resonances acting as doorway states to the continuum states in the vacuum. 
    We are able to quantify the effective coupling strength of these doorway states to the continuum for single-layer graphene,  bilayer graphene, and graphite at the $\Gamma$ point for states between the vacuum energy $E\textsubscript{vac}$ and $E-E\textsubscript{vac}= +15$\,eV. 
    By applying density functional theory (DFT) we show that structures in the DOS and band dispersion alone cannot account for the modulations in the LEE spectrum. 
    Instead, an ensemble of material-specific and layer-number-specific doorway states are key in shaping the structured LEE spectrum.
    As the properties of Feshbach resonances can be traced to their interlayer character, we expect modulations of the LEE spectrum to be a generic phenomenon of layered materials.
    Feshbach resonances result from the 
    energetic degeneracy and coupling between quasi-bound and continuum states.
    Interestingly, this type of resonance occurs in the present case of layered materials and allows to identify Feshbach resonances as doorway states to LEE emission.

    \begin{figure}[b!]
    	\includegraphics{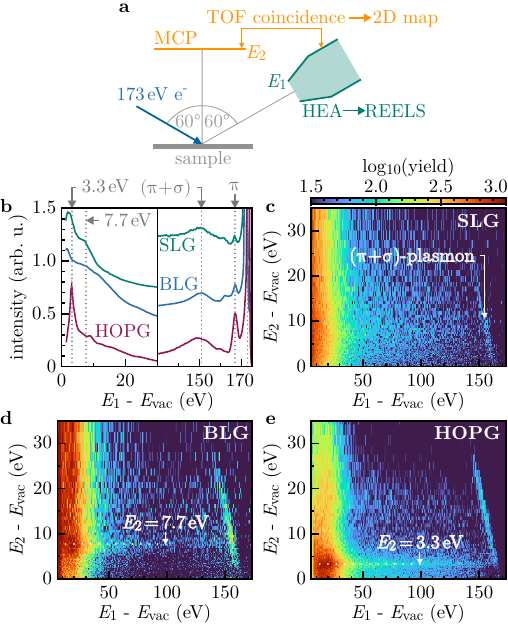}
    	\centering
    	\caption{Low-energy electron spectroscopy. (a)~Schematic of the experimental setup: 173\,eV electrons impact on a sample surface under 60$^\circ$ with respect to the sample surface normal: Either one electron ($E\textsubscript{1}$) or two coincident secondary electrons ($E\textsubscript{1}$,$E\textsubscript{2}$) are detected energy-resolved with a hemispherical energy analyser (HEA, $E\textsubscript{1}$) and a multichannel plate (MCP) detector using a time-of-flight (TOF) approach ($E\textsubscript{2}$), respectively. Reflection electron energy loss spectra (rEELS) are shown for quasi-freestanding single-layer graphene (SLG, green), bilayer graphene (BLG, blue) and highly-oriented pyrolitic graphite (HOPG, red) in panel (b). Corresponding ($E\textsubscript{1}$, $E\textsubscript{2}$) correlation maps for all three materials are shown in panels (c), (d), and (e), respectively. All heat maps use the color scheme indicated on top of panel (c) adapted to represent the logarithm of the respective electron yield counts.}
    	\label{fig:Fig1}
    \end{figure}
    
    In this work we explore the LEE emission from single-layer graphene (SLG) and bilayer graphene (BLG) as well as graphite, the bulk limit of many layers. 
    In the experiments, we use quasi-freestanding layers of SLG and BLG. 
    The samples were fabricated by annealing epitaxially grown zero-layer and monolayer graphene on 4H-SiC(0001) in hydrogen atmosphere at 550$\,^\circ$C and 860$\,^\circ$C, respectively. 
    The graphene used as a precursor was produced by adapting the procedure from Kruskopf~\textit{et~al.}~\cite{kruskopf2016comeback}.
    Additionally, we performed measurements using a highly-oriented pyrolitic graphite (HOPG) sample of ZYB quality with a nominal mosaic spread of $(0.8\pm 0.2)^\circ$ which was mechanically exfoliated in air before being introduced in the vacuum chamber. 
    All samples were annealed under ultra-high vacuum conditions at $500^\circ$C (HOPG) and $350^\circ$C (SLG, BLG) prior to and several times during week-long measurements. 
    The vacuum pressure during the measurement was $\le 4 \times 10^{-10}$\,mbar.
    Fig.~\ref{fig:Fig1}~(a) shows the experimental geometry used to measure electron pair emission. 
    The final energy $E_1$ of scattered 173\,eV electrons, incident at an angle of $60^\circ$ relative to the surface normal of the target, is measured using a hemispherical energy analyser (HEA) at $60^\circ$ with respect to the sample surface normal which corresponds to a Bragg maximum for this geometry. 
    A microchannel plate (MCP) detector with a RoentDek delay line anode is placed above the surface and provides the energy information on the secondary slow electron ($E\textsubscript{2}$) for coincidence measurements using a time-of-flight (TOF) approach (cf.~\cite{werner_secondary_2020}). 
    Despite the indistinguishability of electrons, the identification of scattered primary and secondary electrons in this (e,2e), i.e. one electron in, two electrons out,  process is well justified in the present case for the region of strongly asymmetric energy distribution $E\textsubscript{2} \ll E\textsubscript{1}$ (Fig.~\ref{fig:Fig1}~(b)) we focus on since the exchange amplitude is small compared to the direct scattering amplitude~\cite{taylor2012scattering}.
    Note that all energies are given with respect to the sample vacuum level.
    Low currents of 2.6\,pA were used for coincidence studies to ensure a reasonable true-to-false ratio on the order of unity.     
    Fig.~\ref{fig:Fig1}~(b) shows (non-coincident) reflection electron energy loss spectra (rEELS) for all three materials for a primary beam energy of 173\,eV. 
    While the LEE spectra (left side) look different for the three materials, only the HOPG spectrum displays the pronounced `X-peak' at 3.3\,eV. 
    The right side of Fig.~\ref{fig:Fig1}~(b) displays the energy loss spectrum of the primary electron where the elastic peak (zero energy loss at 173\,eV) has been cut-off to more clearly display the $\pi$- and $(\pi + \sigma$)-plasmon peaks. 
    Smaller shoulders may indicate double and triple plasmon excitation.
   
    In the $(E\textsubscript{1},E\textsubscript{2})$ coincidence maps for SLG, BLG and HOPG measured under identical conditions (Fig.~\ref{fig:Fig1}~(c)-(e)) we can identify an island near $E\textsubscript{1}-E\textsubscript{vac}\sim 150$\,eV stretching from $3$\,eV$\lesssim E\textsubscript{2}-E\textsubscript{vac}\lesssim 30$\,eV, representing the excitation and direct decay of a $(\pi + \sigma)$-plasmon. 
    The $\pi$-plasmon excitation, while visible in panel~(b), is not present in the $(E\textsubscript{1},E\textsubscript{2})$ coincidence maps because, for the present experimental geometry, observation of the direct decay of the $\pi$-plasmon is  kinematically forbidden~\cite{werner_secondary_2020}.

    \begin{figure}[h!]
    	\includegraphics{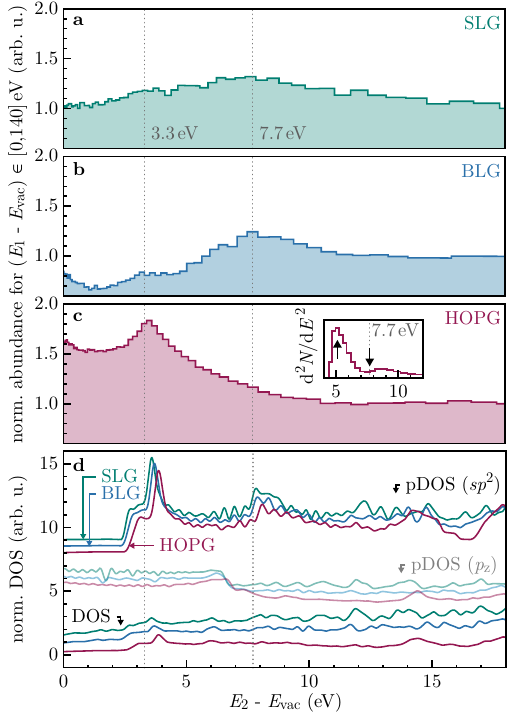}
    	\centering
    	\caption{Secondary low-energy (LEE) spectra for (a)~SLG, (b)~BLG and (c)~HOPG. These $E\textsubscript{2}$ spectra were obtained by integrating the double differential data in Fig.~\ref{fig:Fig1}~(c)-(e) over $E\textsubscript{1}$ in the range $[0,140]$\,eV and normalised to the bin at 18\,eV. The inset for HOPG displays the second derivative of the energy spectrum. (d)~Density of states (DOS) and projected DOS (pDOS) for $sp\textsuperscript{2}$ and $p\textsubscript{z}$ orbitals,  shifted vertically for better visibility.}
    	\label{fig:Fig2}
    \end{figure}
    
    In the region of low-energy secondary electron emission ($E\textsubscript{2}\lesssim 30$\,eV), where in the rEELS spectrum only HOPG exhibits a strong peak at $3.3$\,eV, the $(E\textsubscript{1},E\textsubscript{2})$ map reveals a distinct $E\textsubscript{2}$ feature also for BLG, however, at a different energy of 7.7\,eV. 
    Both features are also clearly visible in the projected $E\textsubscript{2}$ spectra (Fig.~\ref{fig:Fig2}), where we have integrated the joint $(E\textsubscript{1},E\textsubscript{2})$  spectra for all three materials over the energy interval $0\le E\textsubscript{1}-E\textsubscript{vac}\le 140$\,eV. 
    The upper cut-off is chosen to suppress the distortion of the projected spectra by the direct plasmon energy loss peak. 
    For SLG (Fig.~\ref{fig:Fig2}~(a)), the spectrum is largely structureless with only a slight enhancement around 7.7\,eV while BLG (Fig.~\ref{fig:Fig2}~(b)) displays a more pronounced asymmetric peak at this energy and only a small peak at 3.3\,eV which dominates the LEE spectrum of HOPG (Fig.~\ref{fig:Fig2}~(c)). 
    Variation of the $E\textsubscript{1}$ window over which the projection is performed leaves the position of these peaks unchanged (see Appendix~\ref{sec:app_projections}), indicating the absence of significant energy-momentum dispersion of the spectra. 
    This demonstrates even further that the low-energy electron emission from these van-der-Waals materials is not a result of a direct electron-electron scattering but dominated by plasmon decay into electron-hole excitation. 
    
    The plasmon can either directly decay via emission of an electron (cf. the island in the $(E\textsubscript{1},E\textsubscript{2})$ maps in Fig.~\ref{fig:Fig1}) or populate excited states with energies above the vacuum level which eventually decay resonantly into continuum states in the vacuum. 
    The observed LEE spectrum is thus determined by the population of these quasi-bound ($E \ge E\textsubscript{vac}$)
    resonant states and their subsequent coupling to the continuum.

    To elucidate the origin of the distinct features in the LEE spectra, we have calculated the electronic structure of SLG, BLG and HOPG using the Vienna ab-initio software package (VASP) (see Appendix~\ref{sec:app_coupling} for details). 
    Starting point is the DOS and the corresponding projected DOS (pDOS) for the $sp^2$ and $p_z$ orbitals shown in Fig.~\ref{fig:Fig2}~(d).
    Clearly, the density of electronic states provides a first-order estimate of the distribution of accessible energies for LEE emission. 
    Indeed, all three materials feature a pronounced and nearly identical DOS peak at $\sim 3.3$\,eV above vacuum.
    In particular, the pDOS for the $sp^2$ orbitals features peaks at both relevant energies 3.3\,eV and 7.7\,eV, with the first one being more pronounced. 
    In contrast, for the $p\textsubscript{z}$ orbitals the pDOS exhibits only one step at $\sim 7$\,eV, which is not reflected in the present experimental spectra. 
    There are no apparent differences either in the pDOS or in the total DOS for the three materials which indicates that the population of excited states should proceed similarly. 
    
    In view of the close similarity of the pDOS for these three layered materials (Fig.~\ref{fig:Fig2}~(d)), the marked difference in the LEE spectra (Fig.~\ref{fig:Fig2}~(a)-(c)) appears quite surprising and unexpected. 
    To explore its underlying mechanism we have performed a stabilisation analysis \cite{Mandelshtam_1993,Burgdoerfer_1994} of the excited quasi-bound states that contribute to the peak in the pDOS. 
    These band structure states above the vacuum threshold can be viewed as Feshbach resonances coupled to the continuum states in the vacuum. 
    Accordingly, we perform slab calculations consisting of the material itself (SLG, BLG or a large number of layers (14 or 22) of graphene representing HOPG) and a region of vacuum of height $L_{\mathrm{vac}}$  separating the periodic replica of the material in the $z$-direction normal to the surface layer. 
    The finite vacuum region delimited by $L_{\text{vac}}$ supports a discrete representation of the emission continuum. 
    When varying $L_{\text{vac}}$, the energies of the discretised continuum states scale as $\varepsilon_{\text{cont}} \propto 1/L^2_{\text{vac}} >0$  while energies of quasi-bound states localized inside the material with resonance energies $\varepsilon_{\text{R}} > 0$ above the vacuum threshold remain (nearly) independent of $L_{\text{vac}}$. 
    This distinct parametric variation of the spectrum as a function of $L_{\text{vac}}$ not only allows to disentangle band structure from continuum states but also permits to quantify the strength of coupling of the quasi-bound states to the continuum through the size of the avoided crossing between $\varepsilon_{\text{cont}}(L_{\text{vac}})$ and $\varepsilon_{\text{R}}(L_{\text{vac}})$: the larger the avoided crossings, the better the coupling between the continuum states and the respective (resonant) quasi-bound states (see Appendix~\ref{sec:app_coupling}).

    \begin{figure}[h!]
    	\includegraphics{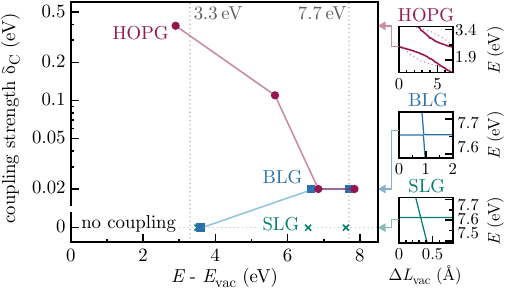}
    	\centering
    	\caption{The logarithm of the magnitude of the coupling matrix element of quasi-bound states above the vacuum level $|\phi_{\mathrm{R}}\rangle$ and continuum states $|\phi_{\mathrm{cont}}\rangle$, $\delta_C$, see Eq.~\ref{eq:defdeltaC}, is shown. Connecting lines to guide the eye. In the panels on the right, we show for each material (each also representing a different coupling strength) a typical (avoided) crossing appearing at the $\Gamma$ point as $L_{\text{vac}}$ is varied. The data from Fig.~\ref{fig:App3} is shown on a magnified scale in dependence of $\Delta L\textsubscript{vac}=L\textsubscript{vac}-L\textsubscript{vac, material}$ with $L\textsubscript{vac, SLG}=14.0$\,\AA, $L\textsubscript{vac, BLG}=17.8$\,\AA, and $L\textsubscript{vac, HOPG}=15.5$\,\AA.}
    	\label{fig:Fig3}
    \end{figure}

\begin{figure}
    \centering
    \includegraphics[]{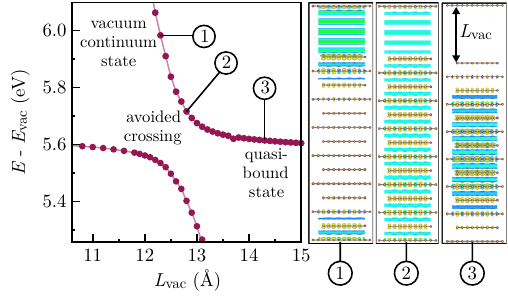}
    \caption{Wave functions of the quasi-bound states and the discretised vacuum continuum states of HOPG (red dots) undergoing an avoided crossing. In the frames \ding{172}-\ding{174} on the right, three exemplary wave functions are shown for an HOPG slab of 14 layers corresponding to the avoided crossing as function of $L_{\mathrm{vac}}$ labelled \ding{172}-\ding{174} in the left panel. Colours represent isosurfaces of increasing charge density (from blue to yellow). Rendered using the software Vesta~\cite{momma_vesta_2011}.}
    \label{fig:FIG4}
\end{figure}

    Fig.~\ref{fig:Fig3} summarises the maximum coupling matrix elements between quasi-bound states above the vacuum level $|\phi_{\mathrm{R}}\rangle$ and continuum states $|\phi_{\mathrm{cont}}\rangle$, $\delta_C$, see Eq.~\ref{eq:defdeltaC}. The three panels on the right show examples of typical (avoided) crossings for each of the three materials. For SLG and BLG we find only very small avoided crossings (i.e. $\delta_C<1$\,meV) near 3.3~eV above vacuum indicating a very weak (or absence of) coupling  of the quasi-bound state to the continuum consistent with the absence of a pronounced `X-peak' in the LEE spectrum. 
    However, for HOPG, there are substantial avoided crossings, again in line with the prominent emission of secondary electrons at this energy in the experiment.
    In Fig.~\ref{fig:FIG4}, we showcase the evolution of associated Kohn-Sham eigenstates along the avoided crossing near 5.7.eV for HOPG: the ``continuum'' states (frame \ding{172}) appear as standing waves in the vacuum region of the slab, the quasi-bound Feshbach resonance is localized between the carbon layers within the bulk (frame \ding{174}). At the avoided crossing, we observe a strong hybridization between the two states (frame \ding{173}). Fig.~\ref{fig:App4} shows similar wave functions along sample (avoided) crossings for SLG and BLG.
    
    These avoided crossings are associated with a class of quasi-bound states of HOPG which are dispersive in the $z$-direction and strongly couple to the vacuum continuum states in line with the Feshbach resonances featuring maximum probability density in between the layers of HOPG (Fig.~\ref{fig:FIG4}). 
    The interlayer characteristics, however, fully develops only a few layers below the surface. 
    These interlayer-centered Feshbach resonances thus form the doorway states for electrons excited into the DOS above vacuum, as precursors to free continuum states outside the material.
    
    We conjecture that it is not a suppression of the excitation probability by scattering, but the suppression of the \textit{coupling} to the vacuum continuum states that is responsible for the shape of the observed LEE spectra (Fig.~\ref{fig:Fig1} and Fig.~\ref{fig:Fig2}). 
    At $7.7$\,eV, avoided crossings for HOPG are smaller by an order of magnitude compared to the ones at 3.3\,eV. 
    Therefore, a pronounced peak at 7.7\,eV is absent although careful analysis, indeed, shows a small feature also at 7.7\,eV in the double-differential spectrum (see Fig.~\ref{fig:Fig1}~(b) and inset of Fig.~\ref{fig:Fig2}~(c)). 
    BLG displays small but finite avoided crossings between quasi-bound states and vacuum continuum states at this energy, consistent with the appearance of a peak in secondary electron emission. 
    Hence the size of the avoided crossing as a function of $L_{\text{vac}}$ and consequently the magnitude of the coupling matrix element between quasi-bound states of the band structure above the vacuum level and plane wave continuum states proves to be key to describe structures of the LEE spectrum. 
    For SLG, the coupling matrix elements for all relevant avoided crossings are $<1$\,meV. 
    For BLG, there is no sizable avoided crossing for the energy level near 3.3\,eV but we find avoided crossings between $6$ and $8$\,eV. For HOPG, avoided crossings appear for all energies with the 3.3\,eV-feature exhibiting a coupling stronger by an order of magnitude than for BLG or SLG.

    The following scenario for LEE emerges: for all three samples, plasmon decay populates the unoccupied band structure above vacuum which does not differ significantly for the three samples of different layer numbers. 
    For SLG, there are no pronounced doorway state favouring the transition to free continuum states which leads to the structureless energy spectrum (Fig.~\ref{fig:Fig2}). 
    In the case of BLG, electrons excited in the bands around 7.7\,eV can escape to vacuum efficiently. 
    Note that the asymmetry seen for the 7.7\,eV-peak for BLG in Fig.~\ref{fig:Fig2} resembles the corresponding peak in the pDOS ($sp\textsuperscript{2}$) indicating that unoccupied states are indeed populated statistically. 
    
    In the pDOS, the most prominent peak for all samples is around $\sim 3.3$\,eV. 
    However, only for HOPG there is a dominant coupling of these interlayer states to vacuum. 
    In addition to the 3.3\,eV-feature, there are also substantial couplings for states at $\sim 5.7$\,eV and $\sim 7-8$\,eV. 
    At these energies, there are no obvious peaks in the $E$\textsubscript{2} spectrum in Fig.~\ref{fig:Fig2}, but the second derivative (cf. insert) shows structures in both cases. 


    The present work highlights the power of coincidence spectroscopy, which allows us to resolve structures in the LEE spectrum otherwise hidden in a broad background of the inelastic secondary electron cascade. 
    In this way, we cannot only identify 
    plasmon-mediated excitation processes but also that there are noteworthy differences in the LEE spectrum shape for different materials. 
    While it is known that the above-vacuum band structure plays a role in the emission process~\cite{PhysRevB.37.4482,strocov_three-dimensional_2000,werner_secondary_2020}, we identify here a second key ingredient for LEE spectra: 
    the coupling strength of these Feshbach resonances to continuum states.
    Our results show the origin of doorway states which open up only above a certain threshold number of layers - the dominant doorway leading to the `X-peak' in HOPG has a threshold of around 5 layers. 

    Our findings provide an important step towards disentangling LEE emission spectra which are rich in information on the surface and its electronic structure. 
    We expect doorway states likely to be key to LEE spectral structures also for other layered materials.

\begin{acknowledgments}
    The authors acknowledge funding from the Austrian Science Fund (FWF) through the grants with DOIs 10.55776/DOC142, 10.55776/COE5 MECS,  10.55776/Y1174, 10.55776/P35539-N and 10.55776/PIN7223324 and support by the Horizon 2022 Marie-Curie Actions Initial Training Network (ITN) EUSpeclab (Grant No. 101073486) funded by the European Union. This project has also received funding from the European Union’s Horizon 2020 research and innovation programme under the Marie Skłodowska-Curie grant agreement No 101022318. The computational results presented have been achieved using the Vienna Scientific Cluster (VSC). 
\end{acknowledgments}

\bibliography{references}

\onecolumngrid
\begin{center}
    \textbf{End Matter}
\end{center}
\twocolumngrid

\appendix

\section{Energy Dispersion of $E\textsubscript{2}$-Peaks}\label{sec:app_projections}

\begin{figure}[h!]
    	\includegraphics{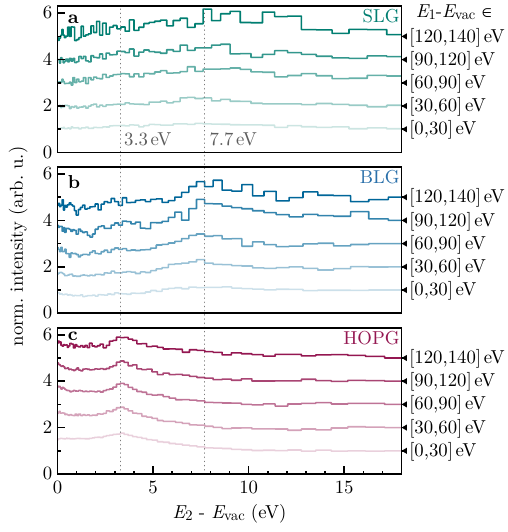}
    	\centering
    	\caption{$E\textsubscript{2}$ energy distribution projections for SLG (a), BLG (b) and HOPG (c). For the five different energy spectra, all counts in Fig.~\ref{fig:Fig1} for a given $E\textsubscript{1}$ energy window labelled on the right-hand side of each panel are summed up. All spectra are normalised to the bin at 18\,eV and shifted along the $y$-axis for better comparison.}
    	\label{fig:App1}
    \end{figure}

    To investigate the dispersion of the horizontal $E\textsubscript{2}$ features in the $(e,2e)$ maps in Fig.~\ref{fig:Fig1}, we have integrated all counts in different $E\textsubscript{1}$ ranges. $E\textsubscript{1}=140$\,eV is considered as maximum energy to prevent an overlap with the $(\pi+\sigma)$-plasmon island. Fig.~\ref{fig:App1} shows these projected $E\textsubscript{2}$ spectra for the single-, bilayer and HOPG sample. For SLG, the spectrum is featureless for all five $E\textsubscript{1}$ ranges. The 7.7\,eV and 3.3\,eV peaks for BLG and HOPG, respectively, are visible in all spectra without significant energy dispersion and changes of the peak shapes.  

\section{Coupling of Local Excitations to Free Continuum States}\label{sec:app_coupling}

    We perform slab calculations of HOPG, BLG and SLG, with periodic boundary conditions in all three spatial directions. 
    We choose the $z$-axis perpendicular to the surface.
    The height of the slab is thus the sum of the height of the material itself, comprised of 1 (SLG), 2 (BLG), and 14 (or 22)  graphene layers for HOPG, and the vacuum $L_{\mathrm{vac}}$ separating  the block of material from its periodic replicas in $z$ direction.
    
\begin{figure}[h!]
    	\includegraphics{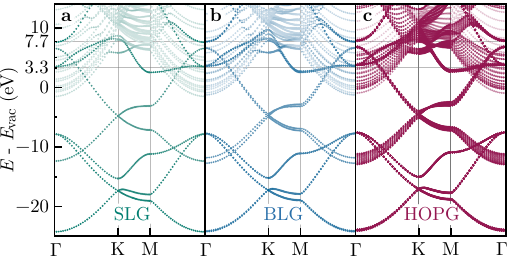}
    	\centering
    	\caption{Band structures of SLG (a), BLG (b) and a 14-layer slab of HOPG (c).
        Discretised continuum states are plotted with slightly transparent color.}
    	\label{fig:App2}
    \end{figure}

    Since we require an accurate model of wave functions above the Fermi level, we employ a large plane-wave energy cutoff of 1200\,eV, and a $k$-point Monkhorst grid of $60\times 60\times 30$ for HOPG and $60 \times 60 \times 1$ for SLG and BLG.
    The supercell for HOPG consists of two layers of carbon atoms in Bernal stacking, forming a hexagonal lattice with a lattice constant of 2.47\,\AA\ and an interlayer spacing of 3.4\,\AA. 
    Above the Fermi level, the associated electronic structure features conduction band states which represent localized resonances in the continuum for energies above the work function of the solid. 
    We consider $L_{\mathrm{vac}} > 10$\,\AA\ to avoid finite-size effects in $z$-direction. 
    Fig.~\ref{fig:App2} shows the resulting band structure of all three materials given with respect to the sample vacuum level for $L_{\mathrm{vac}} =27.2$\,\AA.
    
\begin{figure}[h!]
    	\includegraphics{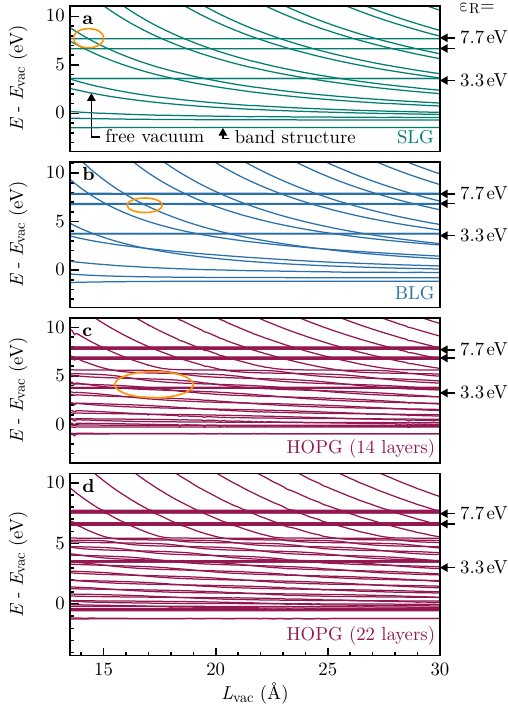}
    	\centering
    	\caption{Energy levels for (a) SLG, (b) BLG, as well as a (c) 14- and a (d) 22-layer slab of HOPG for different simulation cell sizes, consisting of the material slabs plus the vacuum in between $L\textsubscript{vac}$. Energy levels from the band structure at the $\Gamma$-point of the respective material are not influenced by the slab size and can be seen as horizontal lines (see arrows at the right side). Free continuum states decrease in energy with increasing slab size. The crossings of band structure and continuum states (or rather, the avoided crossing) give access to the coupling strength from the material to free vacuum. In (a)-(c), one sample (avoided) crossing is indicated by orange ellipses. These (avoided) crossings are also shown more clearly in the small panels at the right of Fig.~\ref{fig:Fig3}.}
    	\label{fig:App3}
    \end{figure}

    The energy of conduction band (CB) states will only weakly depend on $L_{\mathrm{vac}}$. 
    However, a finite $L_{\mathrm{vac}}$ results in the discretisation of the continuum (cont) states as standing waves in $z$-direction, with $n \lambda \approx 2L_{\mathrm{vac}}$ ($n\in\mathbb{N}$) with $\varepsilon_{\textsubscript{cont}} \propto L_{\textsubscript{vac}}^{-2}$. 
    Consequently, calculating Kohn Sham eigenenergies as a function of $L_{\mathrm{vac}}$ provides a way to disentangle both types of states from each other (Fig.~\ref{fig:App3}). 

\begin{figure} [h!]
    	\includegraphics{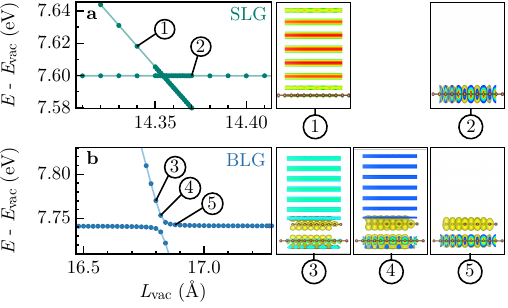}
    	\centering
    	\caption{(Avoided) crossings for SLG (a) and BLG (b). The panels to the right show the corresponding wave functions in the vacuum continuum (\ding{172}, \ding{174}), avoided crossing (\ding{175}, for BLG only) and quasi-bound states (\ding{173},\ding{176}), respectively. The material layers are visualised using grey dots. Colours represent isosurfaces of increasing charge density (from blue to orange). Rendered using the software Vesta~\cite{momma_vesta_2011}.}
    	\label{fig:App4}
    \end{figure}

    round a crossing at $L_{\mathrm{vac}}^0$, the ``diabatic'' conduction-band states $\varepsilon_{\mathrm{CB}}$ appear as (almost) horizontal lines, $\varepsilon_{\mathrm{CB}}(L_{\mathrm{vac}}) \approx \varepsilon^{(0)}$, while the ``diabatic'' continuum energies $\varepsilon_{\mathrm{cont}}$ can locally be well approximated by a first-order Taylor expansion as
    $\varepsilon_{\mathrm{cont}}(L_{\mathrm{vac}}) \approx \varepsilon^{(0)} + \alpha\cdot(L_{\mathrm{vac}}-L_{\mathrm{vac}}^0)$, implying a degeneracy $\varepsilon_{\mathrm{CB}}(L_{\mathrm{vac}}^0)=\varepsilon_{\mathrm{cont}}(L_{\mathrm{vac}}^0)=\varepsilon^{(0)}$. Following degenerate perturbation theory,
    the Hamiltonian around the crossing is then given by
    \begin{equation}\label{eq:defdeltaC}
    H^{(1)} ( L_{\mathrm{vac}})\approx\left(\begin{array}{cc}\varepsilon_{\mathrm{CB}}(L_{\mathrm{vac}})&\delta_C\\\delta_C&\varepsilon_{\mathrm{cont}}(L_{\mathrm{vac}})\end{array}\right).
    \end{equation}
    The eigenenergies of $H^{(1)}({L_{\mathrm{vac}}})$ follow an avoided crossing, with $\delta_C$ the interaction strength  between the conduction-band and the continuum. We extract the interaction strength by fitting $\alpha$ and $\delta_C$ to DFT data.
    Consequently, avoided crossings between vacuum levels with CB states of the same symmetry emerge; some (avoided) crossings are marked with orange ellipses in Fig.~\ref{fig:App3}. Zoom-ins into exemplary crossings of all three materials are shown in Fig.~\ref{fig:Fig3}. Note that we used a smaller $L_{\textsubscript{vac}}$ for the example wavefunctions in Figs.~\ref{fig:FIG4} and \ref{fig:App4} for presentation purposes.
    
    Relating the magnitude of the avoided crossings to the coupling is the key insight underlying the stabilization methods for characterizing Feshbach resonances \cite{Mandelshtam_1993,Burgdoerfer_1994}. 
    The size of these avoided crossings determines the coupling, and thus the interaction strength between the continuum state and the respective conduction band.
    Note that the finite size of the slab supercell results in discretization of both the continuum (represented by only a finite $L_{\textsubscript{vac}}$) as well as the band structure in $z$-direction (represented by a finite number of graphite layers). We have carefully checked both limits numerically: with respect to $L_{\textsubscript{vac}}$ we find convergence to an almost constant $\delta_C$ for $L_{\textsubscript{vac}} > 30$ \AA. Furthermore, the coupling strength per unit energy interval for 14 and 22 layers agree with each other to within about 10\%.
    Examples of the electron density distribution for both cases (and within the crossing) are shown in Fig.~\ref{fig:FIG4} for HOPG and in Fig.~\ref{fig:App4} for SLG and BLG for a characteristic crossing each.


\section{Influence of the Substrate on the Band Structure}\label{sec:app_substrate}

\begin{figure}
    	\includegraphics{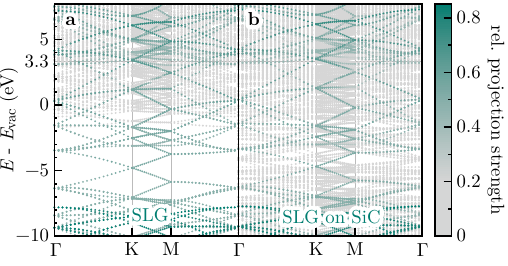}
    	\centering
    	\caption{Comparison of band structure of monolayer graphene~(a) and graphene on SiC with H-intercalation ~(b), in a slab supercell containing $5\times 5$ graphene unit cells, a hydrogen intercalation layer and $4\times 4$ SiC unit cells (with 4 layers in $z$-direction). The color scale shows the relative projection strength $\eta = \sum \left|\left<\phi_{lm}|\psi\right>\right|^2$ of the Kohn-Sham orbitals $\left|\psi_{i\mathbf k}\right>$ on the atomic orbitals $\left|\phi_{lm}\right>$ ($l=s,p,d$) of graphene.}
    	\label{fig:App5}
    \end{figure}

To estimate the potential impact of the SiC substrate, we compare the band structure of monolayer graphene (see main text) with the band structure of a monolayer graphene placed on a SiC substrate with intercalated hydrogen (Fig.~\ref{fig:App5}). For those orbitals located at the graphene layer (green dots), we find only minimal changes and no hybridization with SiC orbitals (grey dots), suggesting that there is no significant impact in the electronic structure of graphene due to the presence of the SiC substrate.

\newpage

\end{document}